\documentclass[
	aps, aps,prd,longbibliography, superscriptaddress, twocolumn,
	10pt
	floatfix, 
    nofootinbib,
	tightenlines
]{revtex4-1}
\usepackage[final]{graphicx}
\usepackage{times,bbm,amsmath,amssymb}
\usepackage{epsfig,color}
\usepackage{xcolor}
\usepackage{hyperref}
\hypersetup{
    colorlinks = true
}
\usepackage{cleveref}
\usepackage{microtype}

\usepackage{float,siunitx}
\usepackage[caption = false]{subfig}

\usepackage[greek,english]{babel}
\usepackage{thumbpdf,enumerate}
\usepackage{booktabs}
\usepackage{sidecap}
\usepackage[scaled=.8]{couriers}
\usepackage{multirow}
\usepackage{placeins}
\usepackage{relsize}
\usepackage{pst-grad,bm}
\usepackage{epigraph}
\usepackage{gensymb}
\usepackage{longtable}
\usepackage{ulem} 
\usepackage{acronym}
\usepackage{physics}
\usepackage{easyReview}

\DeclareUnicodeCharacter{0301}{\'{e}}

\begin{document}

\vspace{10pt}

\title{Effect of Aberrations on 3D optical topologies}

\author{Nazanin Dehghan}
\address{Nexus for Quantum Technologies, University of Ottawa, Ottawa, K1N 6N5, ON, Canada}

\author{Alessio D'Errico} 
\email{aderrico@uottawa.ca}
\address{Nexus for Quantum Technologies, University of Ottawa, Ottawa, K1N 6N5, ON, Canada}

\author{Tareq Jaouni}
\address{Nexus for Quantum Technologies, University of Ottawa, Ottawa, K1N 6N5, ON, Canada}

\author{Ebrahim Karimi}
\address{Nexus for Quantum Technologies, University of Ottawa, Ottawa, K1N 6N5, ON, Canada}
\affiliation{National Research Council of Canada, 100 Sussex Drive, K1A 0R6, Ottawa, ON, Canada}

\begin{abstract}

\end{abstract}

\maketitle

\noindent\textbf{Optical knots and links, consisting of trajectories of phase or polarisation singularities, are intriguing nontrivial three-dimensional topologies. They are theoretically predicted and experimentally observed in paraxial~\cite{berry2001knotted, berry2001knotting,leach2004knotted,dennis2010isolated,leach2005vortex, larocque2018reconstructing} and non-paraxial regimes~\cite{maucher2018creating, sugic2018singular,ferrer2021polychromatic,herrera2022experimental}, as well as in random and speckle fields~\cite{padgett2011knotted}. Framed and nested knots can be employed in security protocols for secret key sharing~\cite{larocque2020optical,kong2022high,ferrer2022secure}, quantum money \cite{farhi2012quantum, aaronson2012quantum}, and topological quantum computation~\cite{bonesteel2005braid}. The topological nature of optical knots suggests that environmental disturbances should not alter their topology; therefore, they may be utilised as a resilient vector of information. Hitherto, the robustness of these nontrivial topologies under typical disturbances encountered in optical experiments has not been investigated. Here, we provide the experimental analysis of the effect of optical phase aberrations on optical knots and links. We demonstrate that Hopf links, trefoil and cinquefoil knots exhibit remarkable robustness under misalignment and phase aberrations. The observed knots are obliterated for high aberration strengths and defining apertures close to the characteristic optical beam size. Our observations recommend employing these photonics topological structures in both classical and quantum information processing in noisy channels where optical modes are strongly affected and not applicable.}

\begin{figure*}
    [!t]
    \centering
    \includegraphics[width=\textwidth]{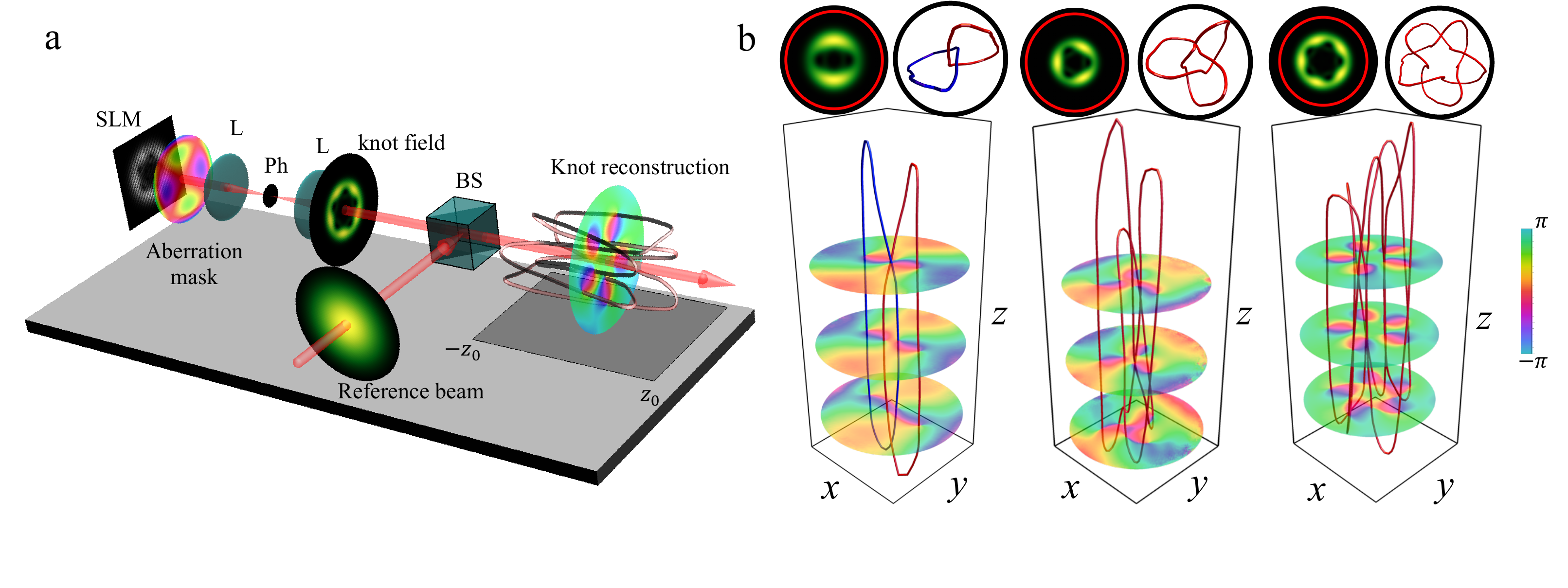}
    \caption{\textbf{Experimental scheme and reconstructed aberrated knots.} {\bf a}- Schematics of the experimental setup. An SLM encodes a phase mask which generates the aberrated knot (Eq. \eqref{eq:aberratedknot}) after a 4-f system and selection of the first diffraction order. The resulting field's phase structure is reconstructed by phase-shifting digital holography, interfering with the knot field and a Gaussian reference beam. Recording the interference patterns at different propagation planes allows one to reconstruct the singular skeleton. {\bf b}- Examples of three-dimensional topologies affected by coma with $\gamma=1$ experimentally reconstructed. From left to right: Hopf link, trefoil, and cinquefoil. The main plots show the singular skeletons and the phase patterns in different propagation planes. Insets show the top view of the singular skeleton (right) and the theoretical intensity patterns in the waist plane with red circles indicating the size of the aperture over which the Zernike polynomials were defined.} 
    \label{fig:fig1}
\end{figure*}
\begin{figure*}
    [!t]
    \centering
\includegraphics[width=\textwidth]{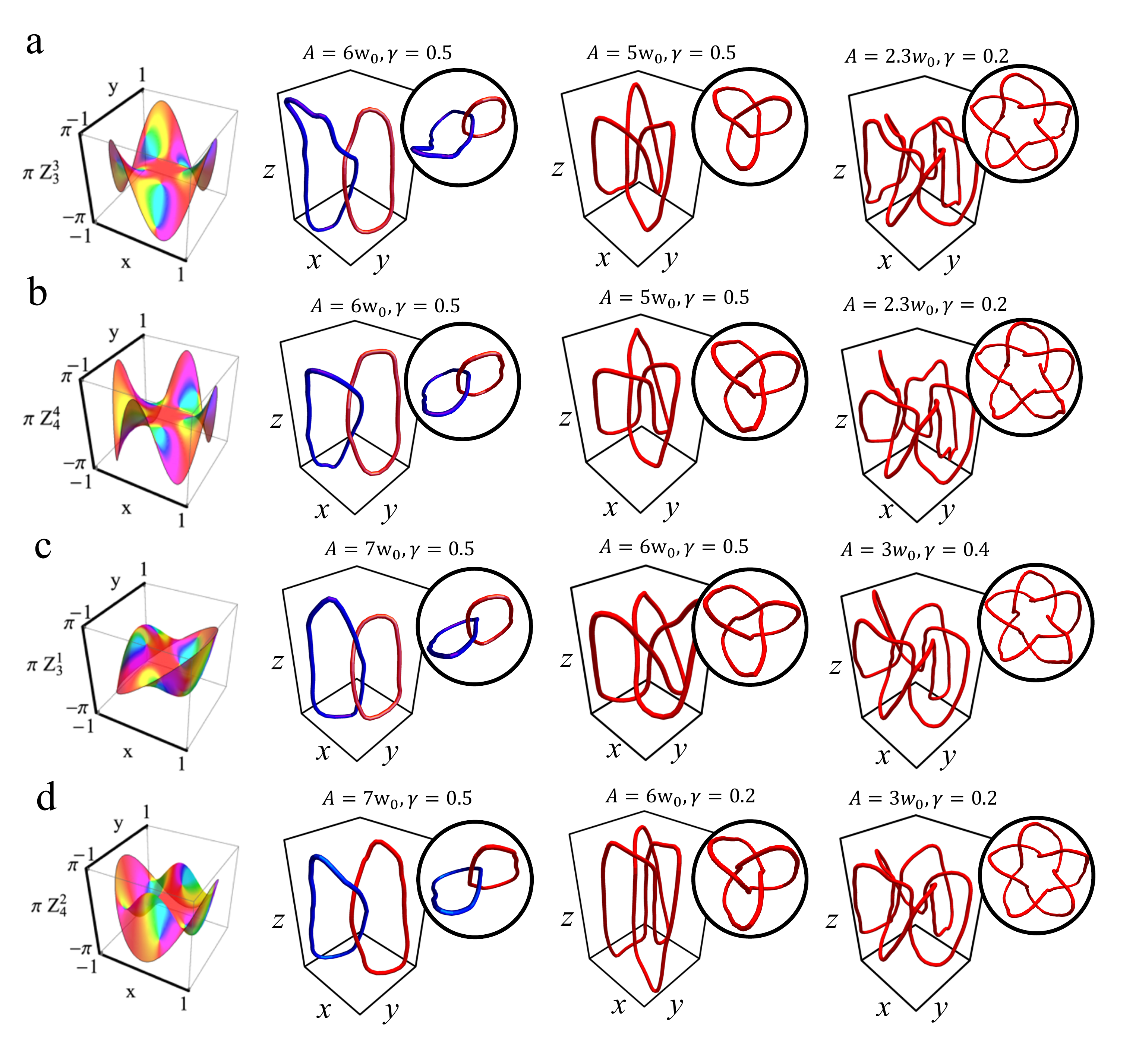}
    \caption{\textbf{Knot survival for small apertures.} The figure shows different singular skeletons obtained for aperture values slightly below the critical aperture  $A_c$ and for strengths $\gamma<1$ such that the topology is conserved. The examples shown are for 3-rd and 4-th order aberrations, namely (a) trefoil, (b) quadrafoil, (c) coma, and (d) secondary astigmatism. The first column shows a three-dimensional plot of the aberrations with a texture given by the cinquefoil phase pattern to highlight how the different Zernike polynomials affect the wavefront near the phase singularities.}
    \label{fig:fig2}
\end{figure*}

Linked or knotted structures can arise from optical fields as trajectories in the three-dimensional space of phase or polarisation singularities~\cite{berry2001knotted,berry2001knotting,dennis2010isolated, dennis2009singular, gbur2016singular}. For instance, a scalar paraxial optical field described by the wavefunction $\psi(x,y,z)$, at a given plane $z$, can exhibit points of zero intensity where the phase is undefined~\cite{nye1974dislocations}. These singular points are characterised by a topological charge $q$ given by the winding of the field's phase around the singularity~\cite{nye1974dislocations, nye1988phase, soskin2001singular,dennis2009singular, gbur2016singular}: $q=1/(2\pi)\oint_{L}\grad{\psi}\cdot d\bm{\ell}$, where $L$ is a closed loop around the singularity and $d\mathbf{\bm{\ell}}$ the infinitesimal arc length. Stable singularities carry the minimal topological charge $q=\pm1$ and can evolve in space with the constraint of total topological charge conservation~\cite{nye1988phase,freund1994wave, freund1999critical, karman1997creation}, i.e. singularity pairs, with individual components having opposite $q$, can annihilate each other or emerge from planes where no charge is present. This dynamics of creation and annihilation of pairs at different planes can result in singularity paths confined in the three-dimensional space and forming a closed trajectory~\cite{freund2000optical}. These curves can be trivial loops, for instance, generated by two charges appearing in one plane and then re-joining each other upon free-space propagation~\cite{berry2006topological}. However, it is now well-known that the wave equation allows for solutions where singularities can track linked or knotted trajectories~\cite{freund2000optical,berry2001knotted}. These solutions were originally found by perturbing high-strength vortex line singularities threaded by unstable loop singularities~\cite{berry2001knotted}. Other approaches based on numerical optimisation procedures and Laguerre-Gauss mode expansions~\cite{leach2004knotted, leach2005vortex, dennis2010isolated, larocque2018reconstructing} have been subsequently developed. More recently, it has been shown that specific optical knots can be generated by imposing weighted polynomials as boundary conditions on the field amplitude~\cite{bode2017knotted,larocque2020optical, kong2022high}. This result, which still lacks a general proof~\cite{bode2017knotted}, is particularly intriguing since it may lead to a systematic approach for generating optical knots and links. This may prove extremely useful in implementing secure communications based on optical knots~\cite{larocque2020optical, kong2022high, ferrer2022secure}, quantum money~\cite{aaronson2012quantum}, and topological quantum computation~\cite{arora2009computational}.

Different knots or links are topologically robust objects since they cannot be smoothly deformed into each other. For a knot to change type, a mathematical transformation should reach a point in parameter space where the knot is singular, i.e. self-intersecting. An optical knot is thus expected to keep the same nature under environmental disturbances that smoothly change its phase and amplitude, thus suggesting a potential advantage for environment-resilient transfer of information. 

However, in practice, typical disturbances may be strong enough to induce a transition in knot type, thus ruling out the potential of this approach. Here we show, through numerical and experimental investigation, how some types of optical knots recently realised experimentally; namely Hopf link, trefoil, and cinquefoil, are remarkably robust under the action of phase distortions applied on the waist plane. The structures mentioned above can all be generated by an optical field which in the plane $z=0$ reads: $\psi(\rho,\phi)=\exp(-\rho^2/(2s^2))\text{Poly}(\rho, \exp(i\phi))$, where $(\rho, \phi, z)$ are cylindrical coordinates, $s$ is a width parameter specifying the size of a Gaussian envelope, and $\text{Poly}(\rho, \exp(i\phi))$ is a Milnor polynomial which specifies the knot type~\cite{bode2017knotted,larocque2020optical,kong2022high} (see Methods for the explicit expressions).

Phase aberrations, which are the most common consequence of environmental disturbances and imperfections in optical setups, can be modelled as an additional phase factor $\delta(\rho,\phi)$ applied on the undistorted field:
\begin{equation}\label{eq:aberratedknot}
    \widetilde{\psi}(\rho,\phi)=\exp(i\delta(\rho,\phi))\psi(\rho,\phi).
\end{equation}
Monochromatic optical aberrations  can be modelled by expanding the phase distortion in Zernike functions:
\begin{equation} \label{eq:zernikesup}
    \delta(\rho,\phi)=\pi\sum_{n,m}\gamma_{n,m}Z_{n}^{m}(\rho/A,\phi),
\end{equation}
where $\gamma_{n,m}$ are real numbers which we call the \textit{strength} of the $(n,m)$-th aberration, and $Z_{n}^{m}(\rho/A,\phi)=R_n^{m}(\rho/A)\cos(m\phi)$ for positive $m$ and $Z_{n}^{m}(\rho/A,\phi)=R_n^{m}(\rho/A)\sin(m\phi)$ for negative $n$, with $R_n^{m}(\rho/A)$ the radial Zernike polynomial, defined in $0\leq\rho\leq A$. We recall that $\abs{m}\leq n$ and $\abs{Z_{n}^{m}(\rho/A,\phi)}\leq 1$.\newline
Here, we mainly investigate the effect of the individual aberrations, up to 4-th order, on optical knots, thus applying phase distortions of the form,
\begin{equation}
    \widetilde{\psi}(\rho,\phi)=e^{i\pi \gamma Z_{n}^{m}(\rho/A,\phi)}\psi(\rho,\phi).
\end{equation}
In particular, we seek the critical values of the strengths $\gamma$ and inverse aperture $1/A$ above which the knot topology is altered. It will be shown how knots can survive these aberrations for apertures slightly larger than the characteristic beam size and strengths $\gamma\approx 1$, which are much higher than values encountered in many practical scenarios. We then push towards smaller apertures to determine which aberrations can affect more the topological structure. It is found that aberrations like coma and secondary astigmatism are the most relevant. This is because these aberrations exhibit local maxima and minima in the interior of the aperture, thus introducing wavefront distortion in between the location of the phase singularities and affecting their trajectories.

\begin{figure}
    [!t]
    \centering
\includegraphics[width=\columnwidth]{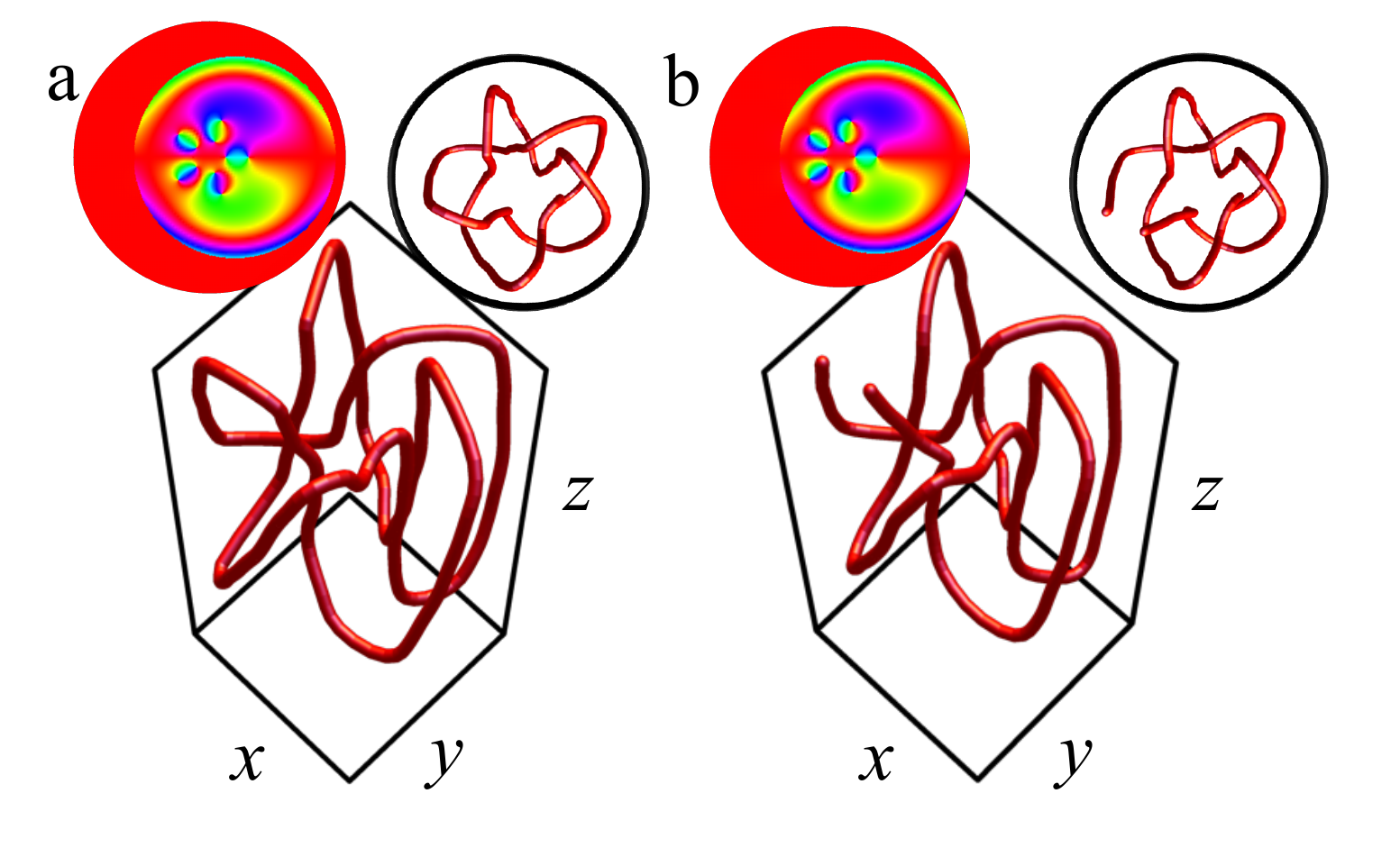}
    \caption{\textbf{Effect of aperture displacement.} The singular skeleton of a cinquefoil knot perturbed by a coma aberration with displaced origin (aperture $A=4 w_0$ and $\gamma=1$). For displacement $\Delta x= w_0$ the knot topology is unaltered (Panel a). For $\Delta x=1.5 w_0$ (b) the singular skeleton opens up due to joining with singularities created in the far field. Insets show the phase pattern in the waist plane (left) and the top view of the singular skeleton (right).}
    \label{fig:fig3}
\end{figure}

\begin{figure}
    [!t]
    \centering
\includegraphics[width=\columnwidth]{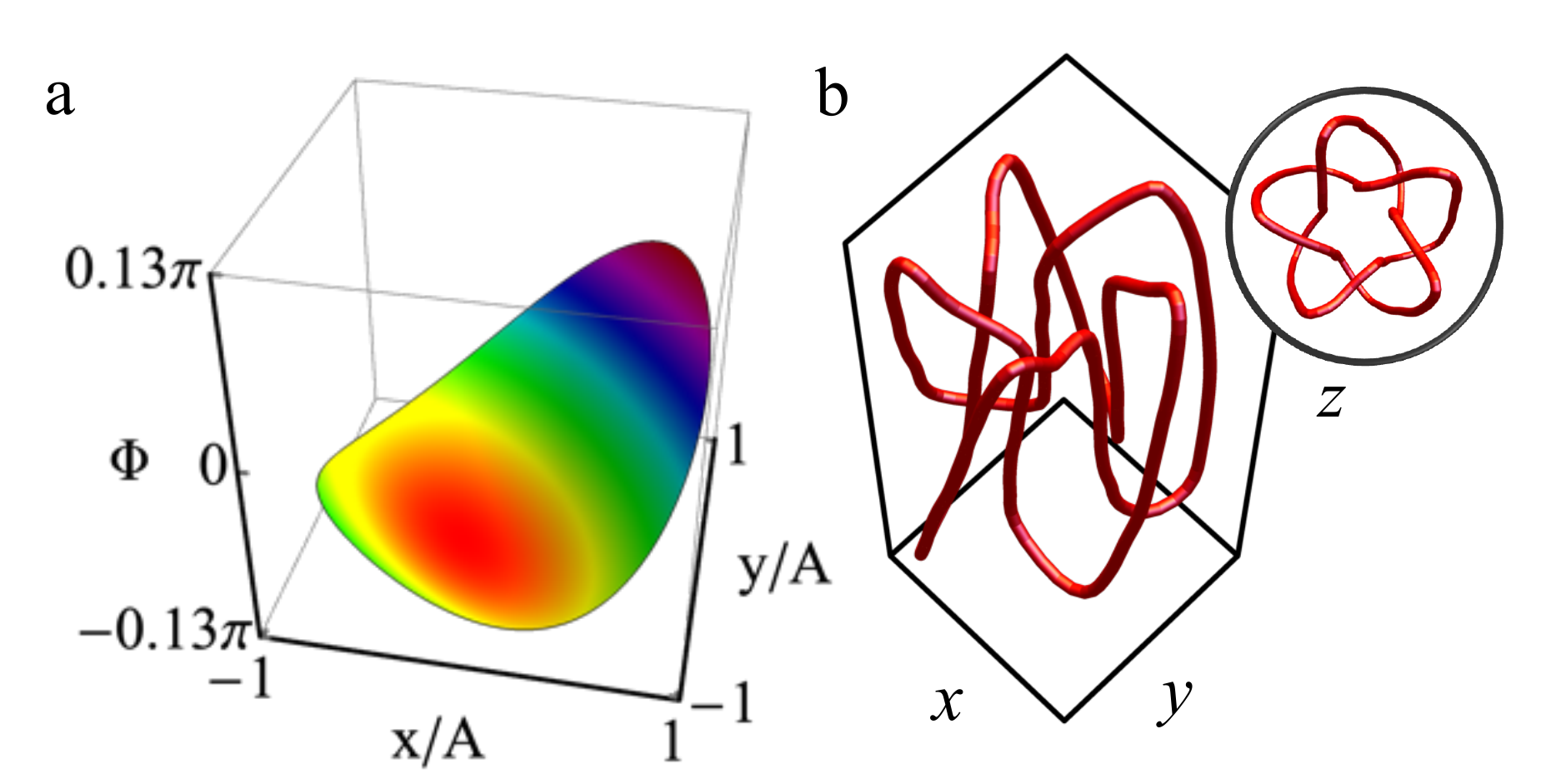}
    \caption{\textbf{Cinquefoil knot subject to a superposition of aberrations.} a- Plot of the wavefront distortion applied to a cinquefoil knot. b- Side view and top view (inset) of the reconstructed singular skeleton.}
    \label{fig:fig4}
\end{figure}
The robustness of three different 3-dimensional optical topologies (Hopf link, Trefoil and Cinquefoil) under various optical phase aberrations was experimentally investigated.

These structures were generated and detected with an approach similar to Ref.~\cite{larocque2018reconstructing}. The principle of the experiment is shown in Fig.~\ref{fig:fig1}-a, while details of the setup are reported in the Methods.  
Optical knots can be obtained from computer-generated holograms displayed on a spatial light modulator (SLM). Exploiting the encoding introduced in Ref. \cite{bolduc2013exact}, a phase mask is applied on an input Gaussian beam which is transformed in the desired knotted field after selecting the first diffraction order (this is done with a pinhole placed in the far field of the SLM). 
In this experiment, the knotted beam interferes collinearly with a reference Gaussian beam. On the translation stage, a CMOS camera is automatically translated to 75 different planes, recording interference patterns that are formed by changing the phase $\alpha$ of the reference beam. The reference phase is controlled using half of the SLM window (see Refs.\cite{larocque2018reconstructing, larocque2020optical} and Methods). The phase of the structured beam in each plane can be reconstructed by means of phase-shifting digital holography. By tracking the phase singularities upon propagation, the singular skeleton was retrieved. Different aberrations, modelled as Zernike functions \cite{born2013principles}, were applied individually as additional phases with a specific aperture on the knot hologram.

In agreement with simulations, it was observed that for each topology and aperture $A$ higher than a critical value $A_c$, the structure survives all different aberrations with $\gamma=1$. For each example, robustness up to at least $\gamma=1$ was observed for values of $A$ slightly larger than the characteristic beam intensity radius. Specifically, the topological structures were unaltered for aperture values $A=8 w_0$, $8 w_0$, and $4 w_0$ for Hopf link, trefoil, and cinquefoil, respectively, where $w_0$ is the characteristic waist parameter (which however, does not necessarily correspond to the beam size, the latter being also dependent on the parameter $s$). We point out that simulations predict that trefoil structures should survive high strength aberrations also for $A=7 w_0$, while experimentally, we observe a critical behaviour for $Z_4^2$ (while other aberrations still give rise to trefoils). We attribute this mismatch to an additional phase perturbation present in the setup. Figure \ref{fig:fig1}-b shows examples of the three structures 
affected by coma (data for the other aberrations are given in the Supplementary Materials, Fig. \ref{fig:aberratedknots}). The main effect of the aberrations is to stretch or compress the individual ``lobes" of the curves leaving the overall topology unchanged. Therefore, if the aperture of the phase aberration compared to the beam waist $w_0$ is bigger than a critical value no aberrations (up to 4-th order) even with $\gamma=1$ can break the topological structure.

When decreasing $A$, knots and links can be broken in open trajectories. This effect can be, for instance, caused by singularity pairs created in the far field, which, instead of annihilating with each other in back-propagation, join with the lines of the singular skeleton. For fixed $\gamma=1$, we define the critical aperture ($A_c$) as the $A$ in which the structure breaks up or cannot be fully reconstructed. The value of $A_c$ is different for different aberrations typologies. For instance, in the case of the cinquefoil, $A_c$ for $Z_{3}^{1}$, $Z_{4}^{0}$, and $Z_{4}^{2}$ is $\approx 3 w_0$ and for $Z_{2}^{0}$, $Z_{2}^{2}$ , $Z_{3}^{3}$, and $Z_{4}^{4}$ is $\approx 2.3 w_0$. When $A \approx A_c$, the effect of $\gamma$ can be further investigated. In Fig.~\ref{fig:fig2}, experimental examples of links and knots under different aberrations with $A\approx A_c$ and $\gamma < 1$ are shown, highlighting how, even if the topology is not preserved under the highest $\gamma$, it can still be recovered for smaller strengths. We note that the experimental knots tend to break up for slightly higher apertures than theoretically expected. This is due to either experimental imperfections, which amount to additional aberrations perturbing the beam, or to residual interference with the zeroth order beam diffracted from the SLM. The latter case was analyzed in detail (see Supplementary Materials) but may be less relevant in practical scenarios where aberrations are applied after the spatial filtering.
\\
The topology of the singular skeleton can be more or less sensitive to different Zernike functions. The aberrations with stronger variations close to the singularities affect their evolution more and, therefore, can destroy the topology for larger $A$ compared with aberrations responsible for distortions of the most external parts of the wavefront. This is the reason why to coma ($Z_{3}^{1}$) and  secondary astigmatism ($Z_{4}^{2}$) is associated a bigger $A_c$ compared to trefoil ($Z_{3}^{3}$) and quadrafoil ($Z_{4}^{4}$) aberrations. As shown in the first column of Fig. \ref{fig:fig2}, coma and secondary astigmatism have local minima and maxima 
in the interior of the defining circle, and so these variations can modify the wave vector distribution around the singularities, thereby altering their trajectory. 
On the other hand, distortions like $Z_{3}^{3}$ and $Z_{4}^{4}$, are mostly flat in their central region while being responsible for wavefront distortions close to the boundary of the defining circle. Hence, for large enough apertures, the phase singularities will lie in the central flat region, and the formation of the singular skeleton will be less affected. 

Up to this point, we considered the case in which the defining aperture of the Zernike functions is perfectly centred with the unperturbed beam. However, lateral misalignments are a common source of imperfections. Hence, we looked into the effect of a relative displacement between the centre of the beam and the centre of the aberration's defining circle. 
Figure~\ref{fig:fig3} shows how displaced coma affects the cinquefoil knot. Relatively small displacements leave the singular skeleton topology unaltered, even for $\gamma=1$. However, for the second displacement ($\Delta x=1.5 w_0$), it can be seen that $Z_{3}^{1}$ breaks the knot. 
We observed qualitatively similar effects for other aberrations (data are reported in the Supplementary Fig. \ref{fig:fig7}).

So far, we have considered high values of the strength by looking at the effect of individual aberrations. In practical scenarios, one deals with wavefront distortions described by a superposition of Zernike polynomials (as in Eq.~\eqref{eq:zernikesup}). As an example, we consider the effect of the wavefront distortion observed in a previous experiment on underwater high-dimensional quantum key distribution (Ref.~\cite{bouchard2018quantum}). The same distortion was applied to the cinquefoil beam, and, as shown in Fig.~\ref{fig:fig4}, the topology survives, which is not surprising since the $\gamma_{n,m}$ are much smaller than 1. We point out that while this wavefront distortion is rather small, it was shown to have a significant effect on the security of OAM-based high-dimensional quantum key distribution. In particular, giving an error rate above the security threshold for dimensions higher than three~\cite{bouchard2018quantum}. This difference in robustness between OAM and knot-based encoding is strictly due to the topological nature of the singular skeletons of structured beams: on the one hand, optical aberrations can abruptly change the decomposition in a given spatial mode basis, thus affecting immediately the fidelity of the transmitted beam and the information encoded within its structure; on the other hand, the change in spatial mode decomposition induced by aberrations does not alter the topological structure associated with a knotted or linked beam, assuming that the strength and/or inverse defining aperture is below a given threshold. Thus, the survival of knot fields under relevant wavefront distortion is a promising example of how these structures can provide a more robust way to encode information. However, we stress that deeper studies on the effects of turbulence must be carried out to certify this advantage. We expect that moderate levels of turbulence, which can be compensated by an adaptive optics system, will not present a serious obstacle to the transmission of either classical or quantum information by means of three-dimensional optical topological structures.

In conclusion, we have demonstrated experimentally how simple optical knots and links are robust under phase aberrations and setup misalignments, thus hinting at their potential advantage in communication protocols. Moreover, we showed how the aberrations exhibiting local minima or maxima in the interior of the circle defining the Zernike polynomials are those which can more significantly affect the topology of the singular skeleton. These considerations can be useful not only for communication purposes but also in devising setups to generate knotted fields in more delicate scenarios, e.g. in nanophotonics experiments~\cite{herrera2022experimental} or in setups for other kinds of structured quantum waves~\cite{harris2015structured, bliokh2023roadmap}, e.g. electrons and neutrons \cite{larocque2018twisting}.


\bibliography{ref.bib}

\vspace{0.5cm}
\vspace{1 EM}

\noindent\textbf{Acknowledgments}
\noindent The authors would like to acknowledge Manuel F. Ferrer-Garcia for providing the first version of the simulation code, and Tugrul Guner for help in the first version of the simulations. This work was supported by Canada Research Chairs (CRC), Canada First Research Excellence Fund (CFREF) Program, NRC-uOttawa Joint Centre for Extreme Quantum Photonics (JCEP) via the High Throughput and Secure Networks Challenge Program at the National Research Council of Canada.  
\vspace{1 EM}

\noindent\textbf{Author Contributions.} N.D., A.D., and E.K. conceived the idea and devised the experiment. N.D., under the supervision of A.D., performed the experiment and analyzed the data. N.D. and T.J. performed the theoretical simulations. E.K. supervised the project. A.D., N.D., and T.J. prepared the first draft of the manuscript. All authors discussed the results and the final version of the manuscript.

\vspace{1 EM}

\noindent\textbf{Data availability}
\noindent
The data that support the findings of this study are available from the corresponding author upon reasonable request.
\vspace{1 EM}

\noindent\textbf{Code availability}
\noindent
The code used for the data analysis is available  from the corresponding author upon reasonable request.

\vspace{1 EM}
\noindent\textbf{Ethics declarations} Competing Interests. The authors declare no competing interests.

\vspace{1 EM}
\noindent\textbf{Corresponding authors}
Correspondence and requests for materials should be addressed to aderrico@uottawa.ca.
\clearpage
%

\section*{Methods}

\subsection{Explicit expressions of the knot fields.}
In the following, we give expressions for the linked and knotted fields in terms of Milnor polynomials modulated by a Gaussian envelope. We chose $a$, $b$ and $s$ in such a way as to obtain the cleanest singular skeleton in the not aberrated case.\newline
The Hopf-link was generated with a mask imposing, on the SLM's image plane, the field amplitude
\begin{align}
    \psi_{\text{Hopf link}}(\rho, \phi)=&e^{-\frac{\rho^2}{2s^2}}\bigr(1-2(1+a^2- b^2)\rho^2 + \rho^4\nonumber\\ &-2(a^2 + b^2) \rho^2 \cos(2 \phi) - 4 i a b\rho^2 \sin(2 \phi)\bigr)
\end{align}
with $a=1, b=1$ and $s=1.7$. Similarly, the trefoil field was generated from the propagation of the amplitude:
\begin{align}
    \psi_{\text{Trefoil}}(\rho, \phi)=&e^{-\frac{\rho^2}{2s^2}}\bigr(-4 \rho ^3 \left(a^2-b^2\right)-2 \rho ^3 (a-b)^2 e^{-i 3 \phi}\nonumber\\&-2 \rho ^3 (a+b)^2 e^{i3\phi}+\rho ^6-\rho ^4-\rho ^2+1\bigr)
\end{align}
with $a=1, b=1$ and $s=1.2$, while the cinquefoil was obtained from 
\begin{align}
    \psi_{\text{cinquefoil}}(\rho, \phi)=&e^{-\frac{\rho^2}{2s^2}} \bigr(1+\rho ^2-2\rho^4-2\rho^6+\rho ^8\rho^{10}\nonumber\\& -16 \rho^5 \bigr(a^2-b^2\bigr)-8 \rho^5 (a-b)^2 e^{-i 5 \phi}\nonumber\\& -8\rho ^5 (a+b)^2 e^{i 5 \phi }\bigr)
\end{align}
with $a=0.5, b=0.4$ and $s=0.65$.

\section{Experimental setup}
\begin{figure}
    [!t]
    \centering
\includegraphics[width=6cm]{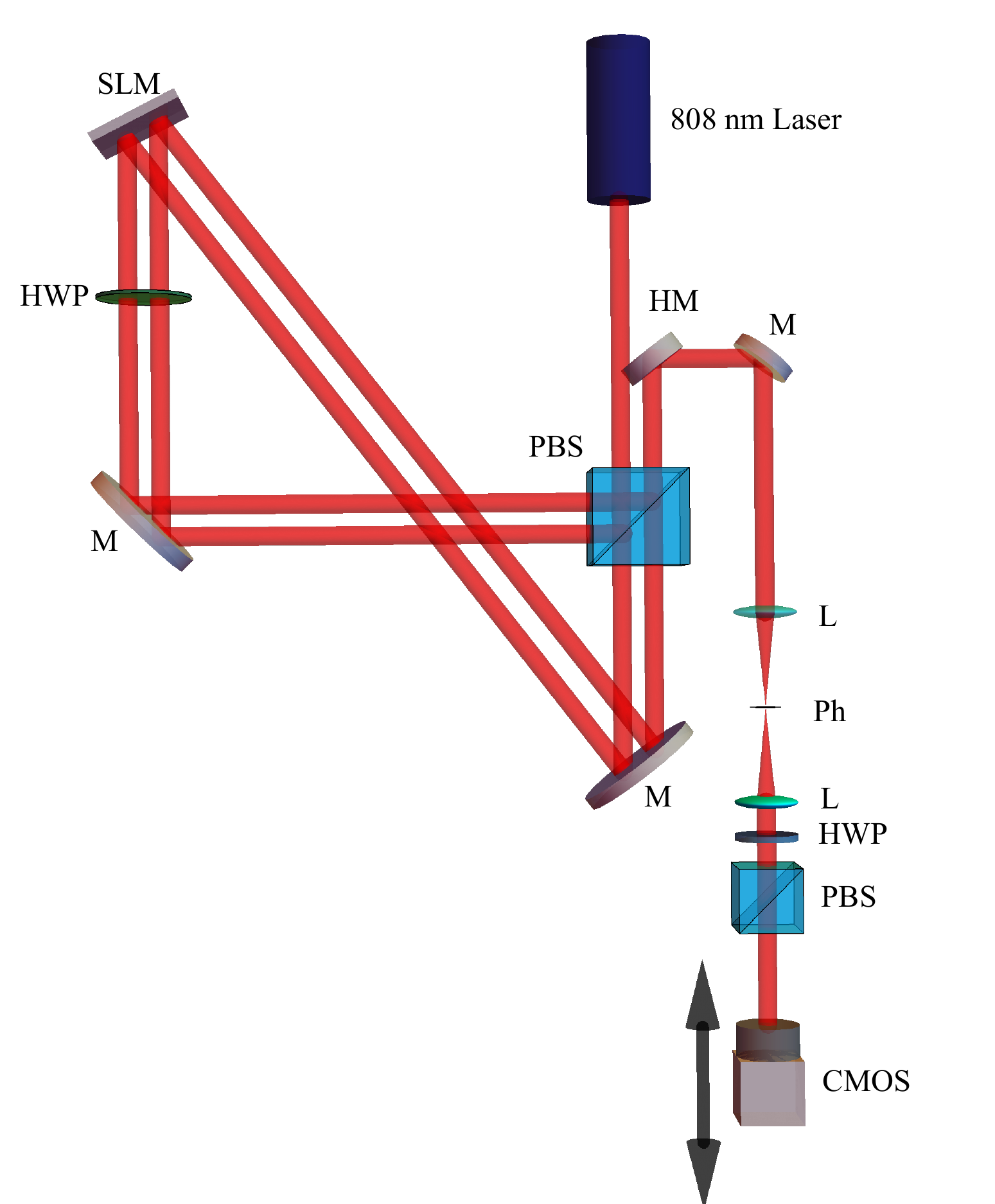}
    \caption{\textbf{Experimental setup} SLM: Spatial Light Modulator, M: Mirror, HWP: Half Wave-Plate, PBS: Polarising Beam-Splitter, L: Lens, HM: D-Shaped Mirror, Ph: Pinhole.}
    \label{fig:figm1}
\end{figure}
In Fig.~\ref{fig:figm1}, we report a more detailed experimental setup. An 808 nm diode laser is coupled to a single-mode fibre, and a Gaussian profile for the beam is generated, which is then sent to a folded-Sagnac interferometer with a Polarising Beam Splitter (PBS) as the input/output port. The vertically polarised beam is converted to a horizontally polarized one by a half-wave plate before reaching the right side of a liquid crystal Spatial Light Modulator (SLM). The SLM replaces the intermediate mirror of the interferometer and displays two holograms on each side, hence applying different transformations on the two beams. On the left side, the knot-generating hologram is applied, while the right side only shifts the phase of the incident beam. Both hologram functions are also superimposed with a blazed grating so that the modulated light is deflected to the first diffraction order. The reference beam is converted to vertical polarization by the same waveplate that acted on the knot hologram's input. In this way, the two beams will both exit the input port of the PBS. The interferometer is aligned to avoid overlap with the input beam and to ensure perfect collinearity of the two output beams. A pinhole placed in the centre of the 4-$f$ system selects the first diffraction order of the SLM. The pinhole aperture and diffraction angle were different depending on the implemented structured beam to minimize the interference with the zeroth order and, at the same time, avoid cutting weak radial contributions necessary for the formation of the desired topology. After the 4-$f$ system the field is $\tilde{\psi}(\rho,\phi,z) \mathbf{e}_{H}+e^{i\alpha}\psi_{\text{ref}}(\rho,z)\mathbf{e}_{V}$ with $\alpha=0,\pi/2,\pi,3\pi/2$, $\tilde{\psi}(\rho,\phi,z)$ the aberrated knot, $\psi_{\text{ref}}(\rho,z)$ the reference Gaussian beam. Projecting on a diagonal polarization (with a half-wave plate and a PBS) we obtained a scalar superposition between the structured beam and the reference. A CMOS camera placed on a motorized translation stage was used to record the intensity patterns of the resulting superposition for different relative phases, thus allowing for phase reconstruction based on phase-shifting digital holography (see next section). 

\subsection{Data analysis.}
The experimental reconstruction of the singular skeleton was obtained by measuring, along the propagation direction $z$, the interference patterns $$I_{\alpha}(z)=\abs{\tilde{\psi}(\rho,\phi,z)+e^{i\alpha}\psi_{\text{ref}}(\rho,z)}^2.$$ From these measurements, the knotted field phase $\xi(\rho,\phi,z)$ (assuming a uniform phase for the reference beam) was retrieved as $$\xi=\text{arg}[I_0-I_{\pi}+i (I_{\pi/2}-I_{3\pi/2})].$$ The coordinates $(x^*,y^*)$ of the phase singularities in each plane were extracted finding the intersections of the lines given by the equations $I_0(x,y)-I_{\pi}(x,y)=0$ and $I_{\pi/2}(x,y)-I_{3\pi/2}(x,y)=0$.
The solution of these equations at different values of $z$ yields a scatterplot of singular points which are then sorted and joined by a path minimization algorithm to reconstruct the singular skeleton.

\clearpage
\onecolumngrid
\renewcommand{\figurename}{\textbf{Figure}}
\setcounter{figure}{0} \renewcommand{\thefigure}{\textbf{S{\arabic{figure}}}}
\setcounter{table}{0} \renewcommand{\thetable}{S\arabic{table}}
\setcounter{section}{0} \renewcommand{\thesection}{Section S\arabic{section}}
\setcounter{equation}{0} \renewcommand{\theequation}{S\arabic{equation}}
\onecolumngrid

\begin{center}
{\Large Supplementary Information for: \\Effect of Aberration on 3D optical topologies}
\end{center}
\vspace{1 EM}

\section{Theoretical simulations}
\begin{figure}
    [!b]
    \centering
\includegraphics[width=15 cm]{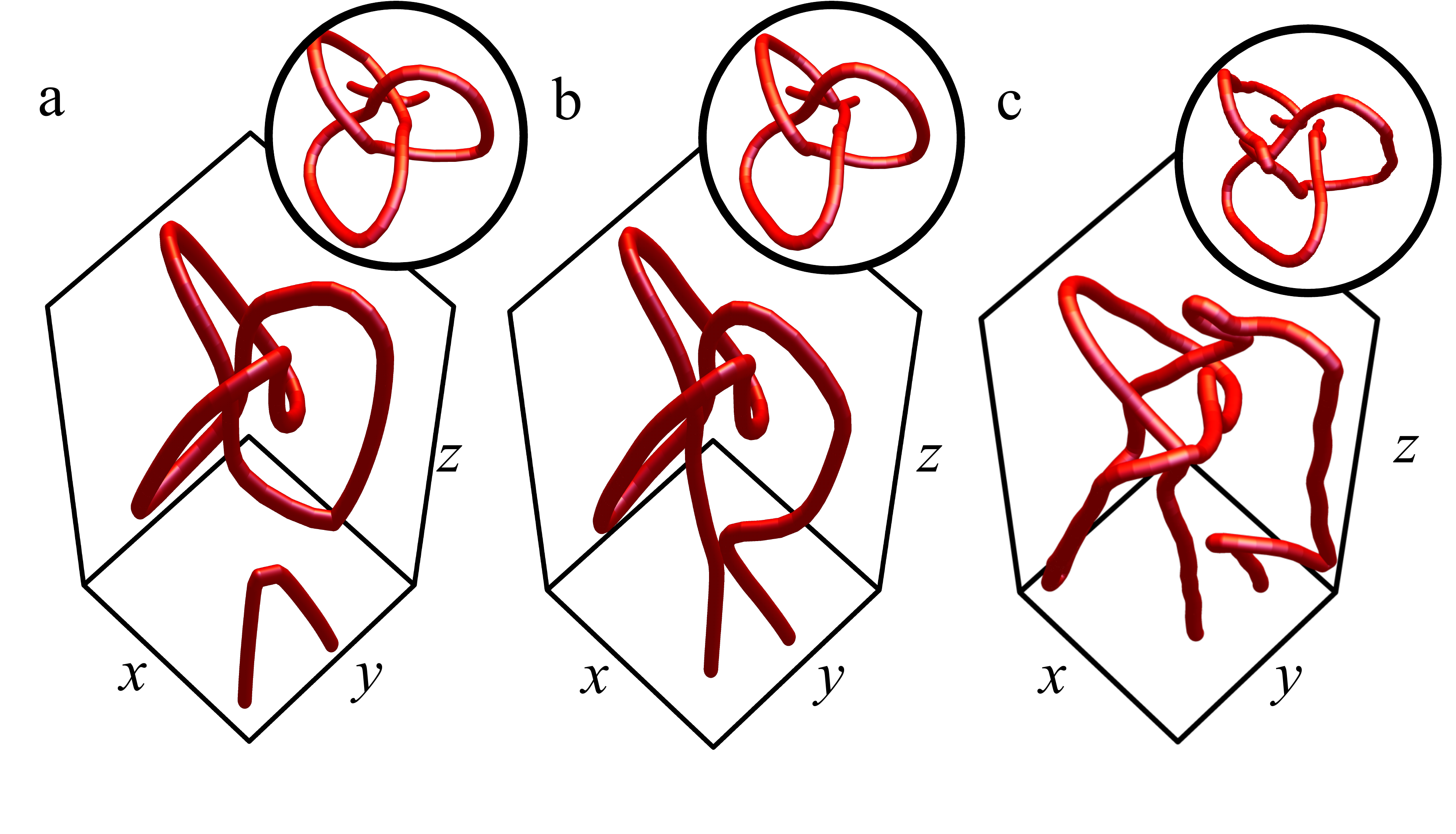}
    \caption{\textbf{Simulating the effects of diffraction from the SLM.} We show the simulation of a trefoil knot affected by secondary astigmatism for aperture $A=6.6 w_0$ without (a) and with (b) the effects of the zeroth diffraction order included. These results partially explain the breakup observed at $A=7 w_0$ in the experiment (panel c). }
    \label{fig:figs1}
\end{figure}

To explore the wide parameter space and choose which case to look for experimentally, we simulated the propagation of the aberrated knotted fields using a Fresnel Transfer Function propagator~\cite{voelzcomputational}. We define each field over a 2048 $\times$ 2048 array and a simulation window with a half-length of $24 w_0$, where $w_0$ is the simulated beam's dimensionless waist parameter, and a numerical wavelength of $0.0134 w_0$. We compute the propagation of the knotted beam given at $z=0$ within the range $z=\pm z_0$, where $z_0=\pi w_0^2/\lambda$ refers to the Rayleigh range. \\
Our experimental results, in some cases, show breakup also in situations where simulations predict the survival of the topological structure. Some discrepancies can be partially explained when considering interference from the zeroth diffraction order of the SLM, which, when applying stronger aberrations, can exhibit a small overlap with the first order. We thus compared the simulations in the ideal case with simulations considering the hologram implementation used in our experiment. We generate a hologram encoding the aberrated knotted field, using the scheme proposed by Ref.~\cite{bolduc2013exact}. The SLM acts as a phase mask applied onto a Gaussian beam with a large  waist (relative to the knotted beam) of $10 w_0$. To simulate the effect of the pinhole placed in the far field of the diffracted beam, we applied a Fourier filter encoded as a Heaviside function $\theta(1 - f_r/\sigma)$, where $f_r^2 = f_x^2 + f_y^2$ is the radial component in Fourier space and $\sigma$ is the pinhole aperture. The value of $\sigma$ was determined experimentally by imaging the iris aperture when illuminated with a Gaussian beam set with waist $w_0$ on the SLM plane. \\
Fig. \ref{fig:figs1}  illustrates an example which demonstrates the effect of simulating the interference of the zeroth order onto the theoretical simulations. When subjected to this aberration at the highest strength and with $A=7w_0$, the  trefoil knot is not observed to survive in the experiment (Fig. \ref{fig:figs1}-c) due to one branch of the singular skeleton opening up. Simulations at $A=7 w_0$ predict, with and without SLM effects included, the existence of the trefoil structure, but with additional singularity lines approaching the knot. Decreasing the value of $A$ to $6.6 w_0$ we indeed observe that, if the zeroth diffraction order is taken into account, the knot breaks up as observed in the experiment (Fig. \ref{fig:figs1}-b) while it does not in ideal conditions (Fig. \ref{fig:figs1}-a). The reason we observe a residual discrepancy in the aperture values must be attributed to additional imperfections in the setup, e.g. deformations of the SLM window itself and imperfect centring of the hologram. 

\section{Supplementary data.}

\begin{figure}
    [!t]
    \centering
\includegraphics[width=\textwidth]{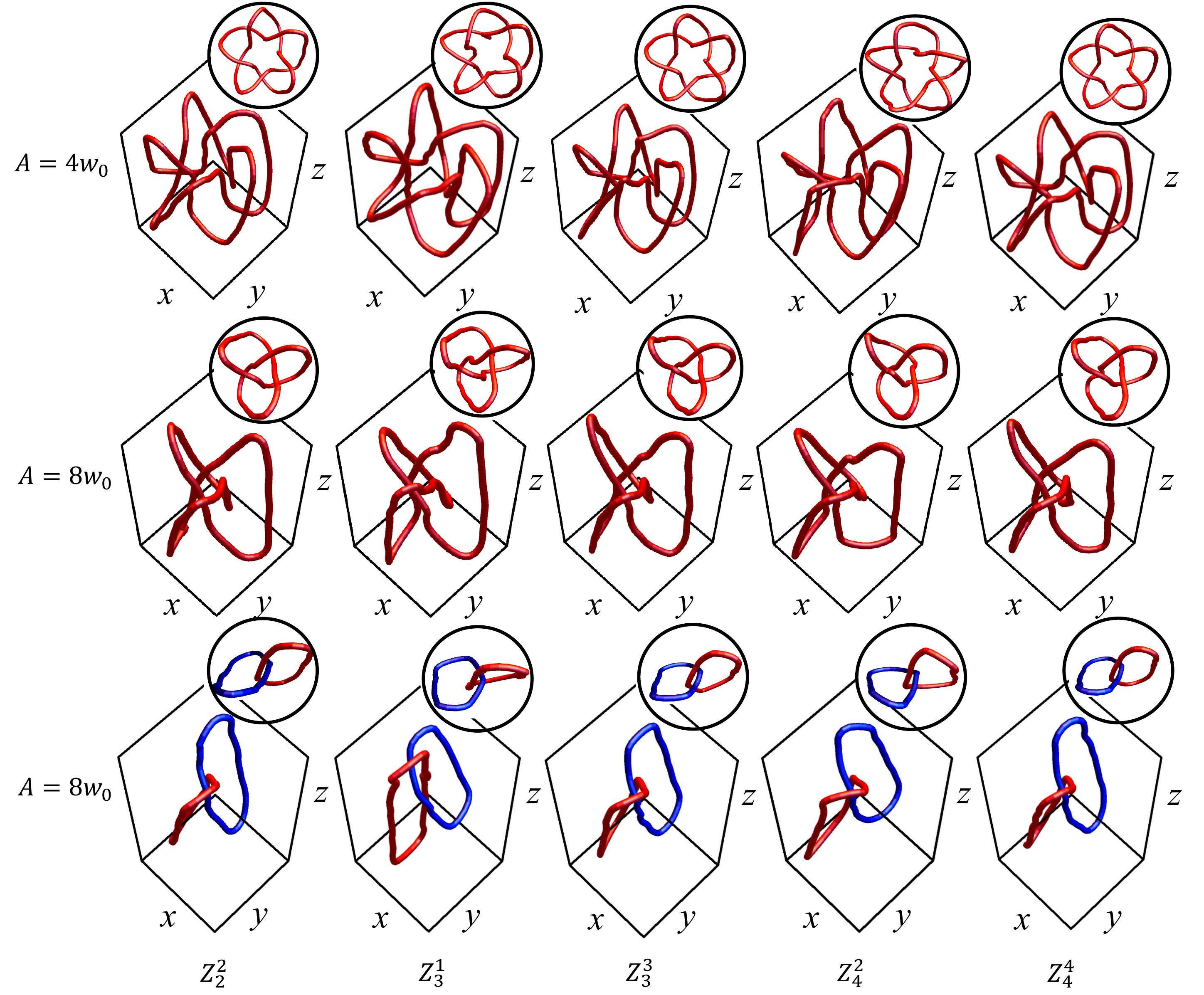}
    \caption{\textbf{Aberrated Knots and links.} Additional examples of topological singular skeleton affected by aberrations with $\gamma=1$ and aperture $A$. The rows represent, from top to bottom, cinquefoils, trefoils and Hopf links, with top views as insets. From left to right, the different singular skeletons were affected by, respectively, astigmatism, coma, trefoil, secondary astigmatism, and quadrafoil aberrations.}
    \label{fig:aberratedknots}
\end{figure}
\begin{figure}
    [!t]
    \centering
\includegraphics[width=\columnwidth]{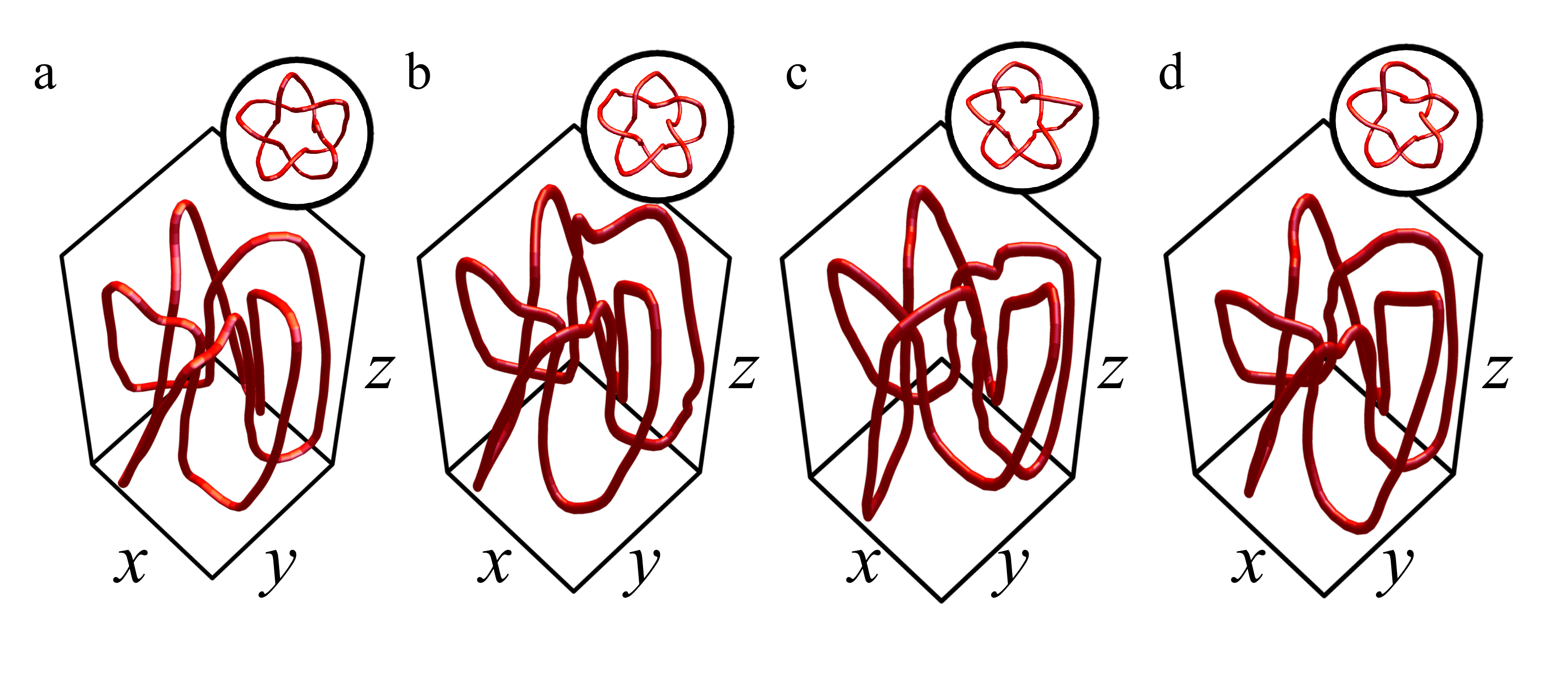}
    \caption{\textbf{Cinquefoil knots under displaced aberrations.} Experimentally reconstructed cinquefoil knots affected by different aberrations displaced by $\Delta x=w_0$ with defining aperture $A=4 w_0$, and strength $\gamma=1$. The different panels correspond respectively to: a, astigmatism $Z_2^2$, b, oblique trefoil $Z_3^3$, c, quadrafoil $Z_4^4$, and d, secondary astigmatism $Z_4^2$. }
    \label{fig:fig7}
\end{figure}
In Fig. \ref{fig:aberratedknots}, we show how for the apertures $A=4 w_0$ (for cinquefoil) and $A=8 w_0$ (for trefoil and Hopf link) the topology of the singular skeletons is not affected, up to $\gamma=1$, by aberrations up to order 4.

Figure \ref{fig:fig7} shows that a displacement of $\Delta x= w_0$ of the window defining the wavefront distortion does not change the topology of cinquefoil knots.

\end{document}